\documentclass{a1}

\title{Glueball propagators in large-$N$ $YM$}

\ShortTitle{Glueball propagators in large-$N$ $YM$}

\author{{Marco Bochicchio} \\INFN-Roma1 and SNS-Pisa \\
Dipartimento di Fisica, Universita' di Roma `Sapienza' \\
Piazzale Aldo Moro 2 , 00185 Roma  \\
       E-mail: \email{marco.bochicchio@roma1.infn.it}}

\abstract{ We have computed in [hep-th/1107.4320] the glueball spectrum in a certain sector of the large-$N$ $YM$ theory by solving by a change of variables
the holomorphic loop equation for cusped twistor Wilson loops supported on certain Lagrangian submanifolds and by evaluating the correlators of surface operators
supported on these Lagrangian submanifolds. 
We have shown that the correlators of composite surface operators of length $L$ 
reproduce in the large-$L$ limit the leading logarithms of perturbation theory of the corresponding glueball propagators, including the correct anomalous dimensions. 
In this paper we show that the correlators of surface operators match in the large-$L$ limit the stronger constraints arising by the operator product expansion,
according to Migdal technique of computing the spectral sum over the glueball including the subleading asymptotics given by the Euler formula.
Finally, we discuss correlators of surface operators for finite $L$. } 

\begin{document}
\def\beq{\begin{equation}}
\def\eeq{\end{equation}}
\def\bea{\begin{eqnarray}}
\def\eea{\end{eqnarray}}
\def\bq{\begin{quote}}
\def\eq{\end{quote}}

\section{Introduction}

The aim of this paper is to show that the recently computed glueball propagators  \cite{MB0} in a certain sector of the large-$N$ $YM$ theory reproduce not
only the logarithms of perturbation theory, including the anomalous dimensions, but satisfy also the more stringent
Migdal criterium \cite{M2} about the operator product expansion ($OPE$). \par
In fact, more precisely, our statement refers to an asymptotic identification for large $L$, discussed in \cite{MB0} and recalled in the following section, between
correlators of composite surface operators of length $L$ supported on certain Lagrangian submanifolds and the corresponding correlators of local composite
single trace gauge invariant operators with momentum dual, in the Fourier sense, to the Lagrangian submanifold. \par
At the end of next section we discuss also correlators of surface operators for finite $L$. \par
Another point of this paper is to discuss the results \cite{MB0} for the glueball propagators in the light of the philosophy of Migdal "meromorphization" \cite{M2}. \par
In this paper we limit ourselves to analyzing the final results without recalling their computation, which is extensively explained in \cite{MB0}. \par
But we recall in this introduction some basic conjectures and results of \cite{M2, M1}. \par
The first basic conjecture by Migdal, that dates long ago \cite{M1}, is that connected two-point correlation functions of local single trace gauge invariant operators, $O(x)$,
of large-$N$ $YM$ or $QCD$ are saturated by a sum of pure poles (for simplicity of notation we consider the scalar case only):
\bea
\int e^{ipx} < O(x) O(0)>_{conn} d^4x= G_O(p^2) = 
\sum_{k} \frac{Z_k(p^2)}{p^2+m_k^2}
\eea
$Z_k(p^2)$ are supposed to be analytic functions of $p^2$ with no poles, positive at the poles of Eq.(1.1), after analytic continuation to Minkowski. \par
The conjecture is based on the estimate that the interaction is suppressed by powers of $\frac{1}{N}$ in the expansion by planar graphs
that defines the large-$N$ limit of $YM$ or $QCD$ \cite{Hooft} and on the assumption that the theory confines the chromoelectric charge, is such a way that the physical spectrum at the leading large-$N$  order is made
by infinite towers of free particles, the glueball and the mesons. \par
In the string picture these particles would arise as the modes of a fluctuating string, open for the mesons and closed for the glueball (see \cite{Pol} and references therein). \par
Naively the two point function, $G_O(p^2)$, has to satisfy the renormalization group ($RG$) equation:
\bea
(\frac{\partial}{\partial \log p}+ \beta(g) \frac{\partial}{\partial g}+2 \gamma_O(g))G_O(p^2)=0
\eea
where $\gamma_O(g)$ is the anomalous dimension
\bea
\gamma_O(g)=\frac{\partial \log Z_O}{\partial \log p}
\eea
and $\beta(g)$ the beta function
\bea
\beta(g)=\frac{\partial g}{\partial \log p}
\eea
More recently it has been argued in \cite{M2} (v.1) that only the spectral function (Minkowski signature is understood):
\bea
\sum_k Z_k(m^2_k) \delta(p^2-m^2_k)
\eea
has in fact to satisfy this property, because of the existence of additive renormalizations of the two-point function of composite operators and
according to the general idea that only the locations of the poles and their residues have a physical meaning.
Therefore it is not restrictive to write \cite{M2}:
\bea
G_O(p^2) = 
\sum_{k} \frac{Z_k}{p^2+m_k^2}
\eea
This is the second basic result of Migdal "meromorphization" \cite{M2}. \par
It is quite clear \cite{M1} why the tower of mesons and glueball has to be infinite, in order to match the perturbative computation of the correlators. \par
Indeed for an operator, $O(x)$, of naive dimension $L$ \footnote{The symbol $\sim$ stays for "equal up to constant irrelevant numerical factors" or "equal up to irrelevant additive terms" depending on the framework.}:
\bea
G_O(p^2) \sim Z^2_O(p^2) p^{2L-4} \log(\frac{p^2}{\mu^2})
\eea
in perturbation theory.
Thus the sum in Eq.(1.6) can reproduce, asymptotically for large $p^2$, Eq.(1.7) only if it is infinite. \par
In perturbation theory the power of momentum, $p^{2L-4}$, arises by naive dimensional analysis. The $\log(\frac{p^2}{\mu^2})$ factor would imply conformal behavior of the correlator at lowest order.
The factor of $Z^2_O(p^2)$ accounts for the anomalous dimension of the operator $O(x)$.
$Z_O(p^2)$ is given at one-loop order in perturbation theory by
\bea
Z_O(p^2)= 1+ \frac{\gamma}{2} g^2 \log(\frac{p^2}{\mu^2})
\eea
for some pure number $\gamma$.
The anomalous dimensions are scheme independent only at one loop. The one-loop $RG$ improved expression for $Z_O(p^2)$ is:
\bea
Z_O(p^2) \sim  \big[\log(\frac{p^2}{\mu^2})\big]^{\frac{\gamma}{2 \beta_0}}
\eea
Recently \footnote{We understand that Migdal argument has been known to him and to his collaborators for long, but never published before \cite{M2}.} it has been pointed out in  \cite{M2} that the spectral sum occurring in the glueball propagator $G_O(p^2)$ has to satisfy, in addition to Eq(1.7), stronger constraints that arise by the $OPE$. \par
The Migdal criterium for the glueball propagators can be described as follows. \par
If the discrete spectral sum on the glueballs is known, it can be evaluated by the Euler formula \cite{M2}:
\bea
G_O(p^2)&&= \sum_k G_k(p^2) \nonumber \\
&&= \int_0^{\infty} G_k(p^2) dk - \sum_{j=1}^{\infty} \frac{B_j}{j!} ( \partial_k)^{j-1}  \big[ G_k(p^2) \big]_{k=0}
\eea
where $B_j$ are the Bernoulli numbers. \par
The leading contribution to the spectral sum arises substituting to the discrete sum the integral.
This has to reproduce the perturbative result, Eq.(1.7). \par
However, Euler formula furnishes the subleading asymtotics to the integral, that contain inverse powers of the momentum $p^2$, because of the derivatives.
These inverse powers are interpreted as non-perturbative contributions arising by condensates of higher dimension operators in the $OPE$ of the Fourier transform of $O(x) O(0)$. \par
In \cite{M2} it is remarked that also these terms in general have to carry anomalous dimensions, otherwise these terms would imply the existence of unknown conserved currents
in $YM$ or in $QCD$, in addition to the traceless part of the stress energy tensor. \par
In this paper we check our results \cite{MB0} for the glueball propagators against Migdal criterium. \par
However, we should mention that there is an alternative school of thought \footnote{We would like to thank Massimo Testa for pointing out to us this different point of view.} that does not give any physical meaning
to the condensates of higher dimension operators, because of the ambiguity associated to additive renormalizations, necessary to make sense of them in perturbation theory. According to this school
of thought the only constraint that can be meaningfully tested in perturbation theory is the leading perturbative term given by Eq.(1.7), the condensates being $0$ to every order of perturbation theory,
because they are proportional to powers of the $RG$ invariant scale:
\bea
\Lambda_{QCD}= \Lambda \exp(-\frac{1}{2\beta_0 g_{QCD}^2}) (\beta_0 g_{QCD}^2)^{-\frac{\beta_1}{2 \beta_0^2}}(1+...)
\eea
the $OPE$ being ambiguous by the aforementioned additive renormalizations and by non-perturbative terms vanishing in perturbation theory.

\section{glueball propagators}

Recently we computed in \cite{MB0} the glueball spectrum in a certain sector of the large $N$ $YM$ theory, by reducing to a critical equation a new kind of loop equation of large-$N$ $YM$ for 
special cusped Wilson loops, called cusped twistor Wilson loops. \par
Twistor Wilson loops are supported on Lagrangian submanifolds of space-time and the correlators that can be computed solving the new loop equation are built by extended objects
supported on these Lagrangian submanifolds, known as surface operators. \par
The crucial property that allows solving the new loop equation for a twistor Wilson loop, as opposed to an ordinary Wilson loop, is the absence of cusp anomaly in the large-$N$ limit for the former. 
This property plays a role in the interpretation of the correlators of surface operators. \par
In fact a whole family of correlators of the Fourier transform of composite operators of naive dimension $4L$, $O^L(p_+, p_-)$, constructed by surface operators supported on the aforementioned Lagrangian submanifolds, are computed in \cite{MB0}:
\bea
&&\sum_k (\frac{\Lambda}{2 \pi})^{8L-8} \hat N'^2 N^2 \rho_k^4   \int <tr_N Tr_{\hat N'}(\mu \bar \mu)^L(x_+, x_-) tr_N Tr_{\hat N'}(\mu \bar \mu)^L(0,0)>_{conn} e^{i(p_+x_-+ p_-x_+)}  d^4x  \nonumber \\
&& \sim <Tr_{\mathcal N} O^L(p_+, p_-)  Tr_{\mathcal N}O^L(-p_+, -p_-)>^{(C)}_{conn}\nonumber \\
\eea
For large $L$ there is an identification \cite{MB0} between the Fourier transform of our composite surface operators and the Fourier transform of composite local operators
\bea
O^L(p_+, p_-) \sim |(F^-_{01}+iF^-_{03})|^{2L}(p_+, p_-,p_+, p_-)
\eea
based on the coincidence of quantum numbers and anomalous dimensions at large $L$ described below. \par
$F^-_{\alpha \beta}$ is the anti-selfdual  ($ASD$) part of the curvature of the gauge connection
\bea
F_{\alpha \beta}^-=F_{\alpha \beta}- \tilde F_{\alpha \beta} \nonumber \\
 \tilde F_{\alpha \beta}= \frac{1}{2} \epsilon_{\alpha \beta \gamma \delta} F_{\alpha \beta}
\eea
$( x_+=x_4+x_1,x_-=x_4-x_1)$ are light-cone coordinates, $(p_+=p_4+p_1, p_-=p_4-p_1)$ are light-cone momenta. \par
 $\Lambda_W$
is the renormalization group invariant scale in the Wilsonian scheme defined below. \par
In four dimensional Euclidean space-time, with coordinates $(z, \bar z, u, \bar u)$, before the analytic continuation to Minkowski, $\rho_k$ is the density, in units of $\Lambda_W^{2}$,
\bea
\rho_k=\sum_p \delta^{(2)} (z-z_p) 
\eea
of surface operators carrying at each lattice point, $p$, magnetic charge $k$ and holonomy valued in the center, $Z_N$, of the gauge group,
i.e. such that:
\bea
e^{2i \mu_p}=e^{i\frac{2 \pi k}{N}}
\eea 
with:
\bea
\mu \sim \frac{1}{2}(F^-_{01}-iF^-_{03}) = \sum_p \mu_p  \delta^{(2)} (z-z_p(u, \bar u))
\eea
and $z_p(u,\bar u)=z_p$.
The dimensionless inverse "string tension" in units of $\Lambda_W^{-2}$ is:
\bea
\alpha'=  \frac{10}{3  \pi} n 
\eea
with the integer, $n \ge 2$, depending on the choice of the renormalization scheme. $n$ can be absorbed in a redefinition of $\Lambda_W$.
The peculiar support, $(x_+, x_-,x_+, x_-)$, of the correlators arises as the projection with Minkowski signature on the base of a Lagrangian submanifold of the twistor space of (complexified) Euclidean space-time that occurs in our approach. \par
The lattice "field of residues", $\mu_p$, and its continuum limit, $\mu(x)$, is dimensionless and normalized in such a way that the correlator in the Wilsonian scheme be renormalization group invariant. \par
In the canonical scheme our result reads:
\bea
&& <Tr_{\mathcal N} O^L(p_+, p_-)  Tr_{\mathcal N}O^L(-p_+, -p_-)>^{(C)}_{conn}\nonumber \\
&& = g^4(- p_+ p_- ) Z^{-\frac{8L-8}{2}}(- p_+ p_-) <Tr_{\mathcal N} O^L(p_+, p_-)  Tr_{\mathcal N}O^L(-p_+, -p_-)>^{(W)}_{conn}  \nonumber \\
\eea 
with :
\bea
&& <Tr_{\mathcal N} O^L(p_+, p_-)  Tr_{\mathcal N}O^L(-p_+, -p_-)>^{(W)}_{conn}  \nonumber \\
&& \sim  \sum_{k=1}^{\infty} \frac{  \Lambda_W^2 k^{2(2L-1)} \Lambda_W^{4(2L-1)}}{ - \alpha'  p_+ p_-+ k \Lambda_W^2} + ...\nonumber \\
&&\sim  \sum_{k=1}^{\infty} \frac{\Lambda_W^2  \big((k  \Lambda_W^2+ \alpha'  p_+ p_-)(k  \Lambda_W^2- \alpha'  p_+ p_-)+(- \alpha'  p_+ p_-)^2 \big)^{2L-1} }{ - \alpha'  p_+ p_-+ k \Lambda_W^2} \nonumber \\
&& \sim  (- p_+ p_-)^{4L-2}  \sum_{k=1}^{\infty} \frac{ \Lambda_W^2}{ - \alpha'  p_+ p_-+ k \Lambda_W^2} + ...\nonumber \\
&& \sim  (- p_+ p_-)^{4L-2} \log \frac{ - p_+ p_-}{ \Lambda_W^2} + ...\nonumber \\
\eea
The dots stand for contact terms, i.e. distributions whose inverse Fourier transform is supported at coinciding points. \par
$g(- p_+ p_-)$ and $Z(- p_+ p_-)$ are the ($RG$ improved) momentum dependent canonical coupling and renormalization factor \cite{MB1}:
\bea
\frac{\partial g_W}{\partial \log \Lambda}=-\beta_0 g_W^3
\eea
and
\bea
\frac{\partial g}{\partial \log \Lambda}=\frac{-\beta_0 g^3+
\frac{1}{(4\pi)^2} g^3 \frac{\partial \log Z}{\partial \log \Lambda} }{1- \frac{4}{(4\pi)^2} g^2 }
\eea
with:
\bea
\beta_0=\frac{1}{(4\pi)^2} \frac{11}{3} \nonumber \\
\eea
where $g=g_{YM}^2 N $ is the 't Hooft canonical coupling constant.  $ \frac{\partial \log Z}{\partial \log \Lambda} $ 
is computed to all orders in the 't Hooft Wilsonian coupling constant, $g_W$, by:
\bea
\frac{\partial \log Z}{\partial \log \Lambda} =\frac{ \frac{1}{(4\pi)^2} \frac{10}{3} g_W^2}{1+cg_W^2}
\eea
with $c$ a scheme dependent arbitrary constant. Eq(2.12) reproduces the correct value of the first and
second perturbative coefficients of the beta function. 
Indeed, since to the lowest order in the canonical
coupling:
\bea
\frac{\partial \log Z}{\partial \log \Lambda}=
\frac{1}{(4\pi)^2} \frac{10}{3} g^2 + ...
\eea
it follows:
\bea
\frac{\partial g}{\partial \log \Lambda}&&=
-\beta_0 g^3+
(\frac{1}{(4\pi)^2} \frac{1}{(4\pi)^2} \frac{10}{3} -\beta_0 \frac{4}{(4\pi)^2} ) g^5 +... \nonumber \\
&&=-\frac{1}{(4\pi)^2}\frac{11}{3} g^3 + \frac{1}{(4\pi)^4} ( \frac{10}{3}
-\frac{44}{3})g^5 +... \nonumber \\
&&=-\frac{1}{(4 \pi)^2} \frac{11}{3} g^3 -\frac{1}{(4 \pi)^4} \frac{34}{3} g^5+...
\eea
The identification occurring in Eq(2.2) is based on the following argument.\par
In the large-$N$ limit there is a sector of the theory that is integrable at one loop \cite{Zar1,Zar2} (and references therein), that is made by operators of $ASD$ or $SD$ type and their covariant derivatives. The corresponding anomalous dimensions can be computed as the eigenvalues of a Hamiltonian spin chain. \par
Indeed, the anomalous dimensions of a number of operators can be computed explicitly solving by the Bethe ansatz the Hamiltonian spin chain in the thermodynamic limit, that corresponds to operators of large
length $L$ and large naive dimension. In particular the anomalous dimensions of the antiferromagnetic ground states of length $L$ (and naive dimension $2L$) turn out to be of the form (see Eq.(27) of \cite{Zar1} and Eq.(5.23) of \cite{Zar2}):
\bea
Z_L= 1 - L g^2\frac{5}{3} \frac{1}{(4 \pi)^2} \log (\frac{\Lambda}{\mu}) +O(L^0)
\eea
The ground state of the spin chain corresponds to the operators with the most negative anomalous dimension, that turn out to be all scalars constructed by certain contractions involving only
the $ASD$ part of the curvature \cite{Zar1, Zar2}. \par
For large $L$ the anomalous dimensions of the composite surface operators $O^L(p_+, p_-)$ \cite{MB0}
agree with the anomalous dimensions of the ground state \cite{Zar1, Zar2} of the Hamiltonian spin chain. \par
The agreement is at one loop 
since anomalous dimensions are universal, i.e. scheme independent, only at one loop. Actually they agree also for $L=1$, 
since in this case the anomalous dimension is determined by the beta function via the factor of $g^4$. \par
The aforementioned asymptotic identification
suggests that all the glueballs in our spectrum are in fact scalar \cite{MB0}, since this is so for the operators that correspond to the ground state of the Hamiltonian spin chain in the thermodynamic limit \cite{Zar2}.  \par
Thus, to summarize, there is an identification between long composite operators built by surface operators and long scalar local operators polynomial in the $ASD$ curvature $F^-_{\alpha \beta}$, that arise
as the ground state of the Hamiltonian spin chain, that characterizes the one-loop large-$N$ integrable sector of $YM$. \par
The identification is based on the coincidence of the anomalous dimensions and quantum numbers, by two completely independent computations. \par
On one side the solution of the ground state of the Hamiltonian spin chain in the thermodynamic limit, i.e. the large-$L$ limit, that involves solving an integral equation for the Bethe ansatz \cite{Zar1, Zar2}. \par
On the other side the computation of the anomalous dimensions of surface operators by combining of the Nicolai map with the localization of the holomorphic loop equation for cusped twistor Wilson loops
on surface operators \cite{MB0}.
In particular the basic anomalous dimension for $\mu_p$ arises by the Jacobian of the Nicolai map, while the scaling with $L$ is a consequence of the localization. \par
The two computations agree in the large-$L$ limit. \par
Now we test our result for the glueball propagators in the large $L$ limit against Migdal criterium. \par
Firstly, we employ the perturbative one-loop $RG$ improved expression for $Z_{O^L}$ in Eq(2.8):
\bea
Z_{O^L}(p^2) &&\sim \big[\log(\frac{p^2}{\Lambda_W^2})\big]^{\frac{\gamma(2L)}{2 \beta_0}} \nonumber \\
\gamma(2L)&&= 2L \frac{1}{(4\pi)^2} \frac{5}{3}
\eea
In evaluating Eq(2.8) we employ the Euler formula, Eq(1.10), for the spectral sum in Eq(2.9).
The leading term, obtained substituting to the spectral sum the integral, reproduces the perturbation theory, Eq.(1.7), as already remarked in \cite{MB0} (see last line in Eq.(2.9)). \par
For the subleading contribution we obtain terms that contain the same overall $Z_{O^L}^2(p^2)$ factor as in Eq(2.8) multiplied by inverse powers of $p^2$,
arising by the derivatives acting in Eq(2.9). 
These terms are interpreted as arising by operators of higher naive dimension, but with the same anomalous dimension, $Z_{O^L}^2(p^2)$, as the ground state of the Hamiltonian spin chain. \par
The existence of such operators is highly non trivial.
In fact in \cite{Zar2} it has been shown that operators constructed by applying $n$ covariant derivatives to the ground state have the anomalous dimensions (see Eq.(6.33) of \cite{Zar2}):
\bea
\gamma(n,L)= - \gamma(L)+ o(\frac{n^2}{L^2})
\eea
Therefore in the large-$L$ limit they are degenerate with the ground state for any finite $n$. QED \par
Secondly, we would like to exploit the definition of the glueball propagators carrying out the meromorphization idea to its last consequences. 
From the point of view of localization of the cusped holomorphic loop equation \cite{MB0} on sectors labelled by the magnetic charge, $k$, it is perfectly consistent to choose the multiplicative renormalization that occurs in
the canonical scheme
at the scale determined by the density of surface operators, $\rho_k$, i.e. on shell from the spectral point of view (see Eq(1.6)). \par
This would match Migdal idea of meromorphization and would agree with the remark that the only physical 
information of the glueball propagator is contained in the spectral density Eq.(1.5).
Therefore in this case, in place of Eq.(2.8), we get:
\bea
&& <Tr_{\mathcal N} O^L(p_+, p_-)  Tr_{\mathcal N}O^L(-p_+, -p_-)>^{(C)}_{conn}  \nonumber \\
&& \sim  \sum_{k=1}^{\infty} \frac{  g_k^4 Z_k^{-\frac{8L-8}{2}} \Lambda_W^2 k^{2(2L-1)} \Lambda_W^{4(2L-1)}}{ - \alpha'  p_+ p_-+ k \Lambda_W^2} + ...\nonumber \\
&&\sim  \sum_{k=1}^{\infty} \frac{\Lambda_W^2   g_k^4 Z_k^{-\frac{8L-8}{2}}\big((k  \Lambda_W^2+ \alpha'  p_+ p_-)(k  \Lambda_W^2- \alpha'  p_+ p_-)+(- \alpha'  p_+ p_-)^2 \big)^{2L-1} }{ - \alpha'  p_+ p_-+ k \Lambda_W^2} \nonumber \\
&& \sim  (- p_+ p_-)^{4L-2}  \sum_{k=1}^{\infty} \frac{ \Lambda_W^2  g^4(k) Z^{-\frac{8L-8}{2}}(k)}{ - \alpha'  p_+ p_-+ k \Lambda_W^2} + ...\nonumber \\
\eea
with
\bea
Z_{k} &&\sim \big[\log(\frac{k}c)\big]^{\frac{\gamma}{2 \beta_0}} \nonumber \\
\gamma &&= \frac{1}{(4\pi)^2} \frac{10}{3} \nonumber \\
g^2_k && \sim \big[\log(\frac{k}c)\big]^{-1}
\eea
and the constant $c$ to be determined by finite parts in the scheme of \cite{MB0}.
Remarkably the same leading perturbative result as in Eq.(2.8)-Eq.(2.9) is obtained:
\bea
&&  (- p_+ p_-)^{4L-2}  \sum_{k=1}^{\infty} \frac{ \Lambda_W^2  g_k^4 Z_k^{-\frac{8L-8}{2}}}{ - \alpha'  p_+ p_-+ k \Lambda_W^2} + ...\nonumber \\
&& \sim  (- p_+ p_-)^{4L-2}   \Lambda_W^2   \int_{1}^{\infty}  \big[\log(\frac{k}{c})\big]^{-2-\frac{(4L-4)\gamma}{2 \beta_0}} d\log({ - \alpha'  p_+ p_-+ k \Lambda_W^2}) + ...\nonumber \\
&& \sim (- p_+ p_-)^{4L-2} \big[\log(\frac{- p_+ p_-}{\Lambda_W^2 })\big]^{-2-\frac{(4L-4)\gamma}{2 \beta_0}+1} +...
\eea
But now in the Euler formula for Eq.(2.19), as opposed to Eq.(2.8), there are not factors logarithmic in momentum in front of the negative powers, $- j$, of $p^2$ in the spectral sum.
Thus these last terms cannot be anymore related, as it has been done before, to the higher dimension condensates, since they lack of anomalous dimensions. \par
Yet, the way out now is that there are overall factors of $(- p_+ p_-)^{4L-2}$ in front of the spectral sum. Thus for every finite $j$ there is always a sufficiently large $L$ for which
the product of the aforementioned factor and of the negative powers is indeed a positive power of $p^2$. But these positive powers are irrelevant contact terms. QED \par 
We would like to remark that within this interpretation of meromorphization it is sufficient that the domain of definition of the $RG$ flow of the coupling constant and of the anomalous dimensions
contains the support of the spectral measure, Eq(1.5), 
without the need of extending the $RG$ flow to zero momentum as in \cite{M2}.
Indeed a basic result of \cite{MB0, MB1}
is that at large-$N$ there is a Wilsonian scheme for which the Wilsonian beta function in the $ASD$ variables is one-loop exact \cite{MB1} and a canonical scheme in which the beta function has a $NSVZ$  form  that reproduces the first two universal perturbative coefficients \cite{MB1} Eq.(2.15). \par
This is typical of localization \cite{MB0} and because of the same reason occurs also in the $\cal{N}$ $=1$  $SUSY$ $YM$ case \cite{NSVZ, Shif},
since in both cases the Wilsonian beta function is one-loop exact, and thus there is a Landau pole in the infrared for the Wilsonian coupling
and, as a consequence, a finite range of the flow for the canonical coupling. \par
But this is not a problem in Eq.(2.19) as opposed to Eq(2.8), since the $RG$ flow needs to be defined only on the support of the spectral sum if we assume meromorphization. \par
Finally, we discuss the interpretation, according to Migdal criterium, of correlators of surface operators for finite $L$. These are purely non-perturbative objects, because
in general their anomalous dimensions do not match anomalous dimensions of any local operator. \par
Now, surface operators saturate at non-perturbative level \cite{MB0} the v.e.v. of twistor Wilson loops, that have no cusp anomaly for large $N$. 
It is known that for ordinary Wilson loops cusp anomaly can be related to the anomalous dimension of twist-two operators (see \cite{K} and references therein).
Therefore the absence of cusp anomaly of twistor Wilson loops can be related to vanishing of anomalous dimensions of certain operators occurring in the definition of
twistor Wilson loops. In fact the v.e.v. of these operators may actually vanish in perturbation theory because of large-$N$ triviality of twistor Wilson loops, due to extensive cancellations
to every order of perturbation theory \cite{MB0}. But these vanishing operators with no anomalous dimension may get non-perturbative contributions from surface operators. \par
Basically the non-perturbative operators with no anomalous dimension associated to surface operators are powers of their lattice density, $\rho_k$, defined in Eq.(2.4). \par
It is only at large $L$ that they decouple from the $OPE$ implied by Eq.(2.19) to match the $OPE$ of the local operators that occur in the ground state of the Hamiltonian spin chain
in the thermodynamic limit. 
Nevertheless, as a matter of fact, the glueball spectrum that is determined by the poles of the correlators of composite surface operators is $L$-independent and stable in the large-$L$ limit.

\section{Acknowledgments}

We would like to thank Alexander Migdal for many clarifying discussions about glueball propagators.
We would like to thank Massimo Testa for a discussion about the meaning of non-perturbative contributions in the $OPE$.

\end{document}